\documentclass[
nofootinbib,
reprint,
superscriptaddress,
showpacs,
amsmath,amssymb,
aps,
prl,
longbibliography,
dvipsnames,
beamer,
]{revtex4-2}

\usepackage{graphicx}
\usepackage{dcolumn}
\usepackage{bm}
\usepackage{filecontents}
\usepackage{amsmath}
\usepackage{mathtools}
\usepackage{array}
\usepackage{tabularx}

\usepackage{textcomp}
\usepackage{siunitx}

\usepackage{xcolor}
\usepackage{ulem}

\begin{document}

\preprint{}

\date{}
 
\title{Exploiting ionization dynamics in the nitrogen vacancy center for rapid, high-contrast spin and charge state initialization}
\author{D. Wirtitsch}
\affiliation{Vienna Center for Quantum Science and Technology, Department of Physics, University of Vienna, Boltzmanngasse 5, 1090 Vienna, Austria}
\affiliation{Austrian Academy of Sciences, Institute for Quantum Optics and Quantum Information (IQOQI) Vienna, Boltzmanngasse 3, A-1090 Vienna, Austria}
\author{G. Wachter}
\affiliation{Vienna Center for Quantum Science and Technology, Department of Physics, University of Vienna, Boltzmanngasse 5, 1090 Vienna, Austria}
\author{S. Reisenbauer}
\affiliation{Vienna Center for Quantum Science and Technology, Department of Physics, University of Vienna, Boltzmanngasse 5, 1090 Vienna, Austria}
\affiliation{AIT Austrian Institute of Technology GmbH, Giefinggasse 4, 1210 Wien}
\author{M. Gulka}
\affiliation{University of Hasselt, Hasselt, Belgium}
\affiliation{Institute of Organic Chemistry and Biochemistry, Academy of Sciences of the Czech Republic, Flemingovo nam. 2, 166 10 Prague 6, Czech Republic}
\author{V. Iv\'ady}
\affiliation{Department of Physics of Complex Systems, ELTE E\"otv\"os Lor\'and University, Egyetem t\'er 1-3, 1053 Budapest, Hungary
}
\affiliation{MTA–ELTE Lend\"ulet "Momentum" NewQubit Research Group
}
\affiliation{Department of Physics, Chemistry and Biology, Link\"oping University, 581 83 Link\"oping, Sweden}
\author{F. Jelezko}
\affiliation{Institute of Quantum Optics, Ulm University, Ulm, 89081, Germany}
\author{A. Gali}
\affiliation{Institute for Solid State Physics and Optics, Wigner Research Centre for Physics, P.O. Box 49, H-1525 Budapest, Hungary}
\affiliation{Department of Atomic Physics, Institute of Physics, Budapest University of Technology and Economics,  M\H{u}egyetem rakpart 3., H-1111, Budapest, Hungary}
\author{M. Nesladek}
\affiliation{University of Hasselt, Hasselt, Belgium}
\affiliation{IMOMEC, Hasselt, Belgium}
\author{M. Trupke}
\email{All correspondence should be addressed to michael.trupke@univie.ac.at}
\affiliation{Vienna Center for Quantum Science and Technology, Department of Physics, University of Vienna, Boltzmanngasse 5, 1090 Vienna, Austria}
\affiliation{Austrian Academy of Sciences, Institute for Quantum Optics and Quantum Information (IQOQI) Vienna, Boltzmanngasse 3, A-1090 Vienna, Austria}

\date{\today}

\textcolor{RedOrange}{}

\begin{abstract}
	We propose and experimentally demonstrate a method to strongly increase the sensitivity of spin measurements on nitrogen-vacancy (NV) centers in diamond, which can be readily implemented in existing quantum sensing experiments. While charge state transitions of this defect are generally considered a parasitic effect to be avoided, we show here that these can be used to significantly increase the NV center's spin contrast, a key quantity for high sensitivity magnetometry and high fidelity state readout. The protocol consists of a two-step procedure, in which the charge state of the defect is first purified by a strong laser pulse, followed by weak illumination to obtain high spin polarization. We observe a relative improvement of the readout contrast by 17$\,\%$, and infer a reduction of the initialization error of more than 50\,\%. The contrast enhancement is accompanied by a beneficial increase of the readout signal. For long sequence durations, typically encountered in high-resolution magnetometry, a measurement speedup by a factor of $>$1.5 is extracted, and we find that the technique is beneficial for sequences of any duration. Additionally, our findings give detailed insight into the charge and spin polarization dynamics of the NV center, and provide actionable insights for direct optical, spin-to-charge, and electrical readout of solid-state spin centres.
\end{abstract}

\maketitle

\section{Introduction}

Spin centers in crystals such as diamond and silicon carbide are prime candidates for the development of quantum sensors given their long quantum coherence lifetimes and strong optical transitions \cite{Doherty2013a,Rondin2014,Kucsko2013,Dolde2011,Neumann2013,Rogers2014,Becker2016,Sipahigil2016,Christle2015,Widmann2015,Fuchs2015,Bosma2018,Wolfowicz2021}. In diamond, the nitrogen vacancy (NV) center is particularly prominent due to its excellent room-temperature spin coherence, high brightness and large optical spin contrast. The defect, which consists of a substitutional nitrogen atom and an adjacent vacancy in the diamond lattice, displays multiple charge states \cite{Pfender2017}, of which the negative state NV$^-$ is commonly discussed due to its optically addressable spin-1 ground state. However, the charge state dynamics have gained increasing interest in recent years, spurred in part by the observation of spin-to-charge conversion \cite{Shields2015}. 
This mechanism forms the basis of photoelectric detection of the NV center's magnetic resonance (PDMR) and its coherent dynamics, down to the single-defect level \cite{Gulka2017,Bourgeois2015a,Siyushev2019a,Gulka2021}. Since this development removes the need for collection optics and single-photon detectors, it makes the NV center system more amenable to integrated technological applications, particularly in compact diamond quantum sensors \cite{sturner2019,kim2019cmos,webb2019}.

The sensitivity of such a diamond sensor depends on the optical spin readout contrast and on the signal strength (i.e. the square root of the photon collection rate), motivating numerous efforts to improve each of these quantities \cite{Barry2020}. Typical single-NV center experiments obtain a readout contrast of about $30\,\%$.

Recently a relative increase of about $10\,\%$ was reported by using a multi-pulse spin initialization routine with short green laser pulses at a single power, applied in alteration with wait periods over a total duration of several microseconds \cite{Song2020}. However, the charge state dynamics of the center were not considered therein.

Charge state initialization is a key factor towards improving the readout contrast, since population in the neutral state (NV$^0$) leads to undesirable background luminescence and reduced brightness. However, a major hindrance in charge state initialization lies in the unfavorable ionization dynamics that only allow unidirectional conversion from NV$^-$ to NV$^0$ via excitation with a wavelength longer than the zero-phonon line of NV$^0$. Unfortunately, no such mechanism has been shown for the transfer from NV$^0$ to NV$^-$. The importance of charge state purity for increased measurement sensitivity was highlighted in another recent experiment, where the center's charge state was controlled in real time via fast electronic feedback \cite{Hopper2020}. Although this method achieved high fidelity charge state initialization and an increase in the contrast by a factor $\sim 1.13$, it required an initialization duration of several tens of microseconds and considerable hardware overhead.

Here, we predict and demonstrate a method to improve the readout contrast and brightness by enhanced charge and spin state initialization of the NV center. The method is readily suitable for existing experimental setups. We find that this sequence provides superior initialization for any measurement duration and we thereby improve the collected photon rates by $11\,\%$ and the maximal readout contrast to $>46\,\%$. It requires only excitation using a single wavelength with different powers, and we demonstrate experimentally that most of the resulting improvement can be obtained using very short pulse sequences with a combined length of less than \SI{3}{\micro\second}, providing enhanced readout for any measurement sequence length. Furthermore, the principle is most likely applicable to a whole host of similar systems, such as the divacancy, the silicon vacancy, and NV centres in silicon carbide, as well as similar defects in other materials \cite{Christle2015,Widmann2015,Fuchs2015,Mu20,Wang21,Wolfowicz2021}. 

\begin{figure}[!t]
	\centering
	\includegraphics[width=\columnwidth]{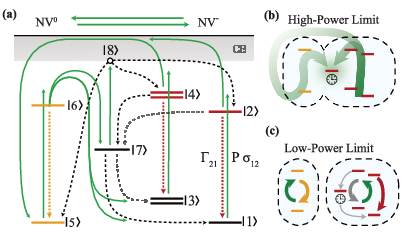}
	\caption{Model Overview. \textbf{(a)} Level diagram (levels 1 to 8) used for model predictions including a charge capture channel into the metastable state  (level 7). Green arrows denoted by a $\sigma$ indicate a laser driven transition while system-internal decay processes are indicated by dashed lines and denoted by a $\Gamma$. \textbf{(b)} Sketch of population distribution within the NV center under high-power illumination. At strong saturation, a significant portion of the NV center population will be shelved in the longest-lived state, i.e. the NV$^-$ metastable state. \textbf{(c)} Under low-power illumination, population within the NV$^-$ manifold will be spin-polarized by the inter-system crossing dynamics, while the quadratic power dependence suppresses two-photon charge state mixing processes.}
	\label{fig:steady_state}
\end{figure}

\section{Model: Charge State Dynamics to Improve Signal and Contrast}

We develop an effective rate equation model which includes both the spin and the charge dynamics to reproduce the experimental time traces and the derived spin contrasts. A conceptual overview of the scheme is shown in Fig \ref{fig:steady_state}a. The model utilizes the decay rates ($\Gamma_{ji}$) and absorption cross sections ($\sigma_{ij}$), with $i$ and $j$ corresponding to the involved levels. The free parameters in the model are the excitation cross-sections, all decay rates, and overall scaling factors for the excitation rates of the two lasers. The resulting parameters are found by a least-squares minimization procedure described in detail in Appendices A, B, and G. Our model consists of five levels in the negatively charged center (two ground states, two excited states and one singlet level) and two levels in the neutral state.
We do not include any metastable levels in the neutral charge state. It is known that the NV$^0$ charge state possesses a metastable manifold which has been observed by electron spin resonance \cite{felton2008electron}. From this manifold a one-photon charge state conversion is energetically possible \cite{Alkauskas2021}. Previous observations, however, indicate that this process does not play a significant role in the charge state dynamics of the system \cite{Aslam2013, baier2020orbital, roberts2019spin}. Finally, a weakly bound state is included as described below.

A two-photon process via level 6 effects the NV$^0$ to NV$^-$ transition \cite{Aslam2013, baier2020orbital, roberts2019spin}. \textit{Ab-initio} considerations indicate that the conversion from NV$^0$ to NV$^-$ occurs from the optically excited NV$^0$ state (level 6) both directly to levels 1 and 3, as well as via the metastable singlet state (level 7) (see Appendix A). This state exhibits a lifetime which is more than an order of magnitude longer than all optically excited states, and has a small excitation cross-section that allows population to be shelved during illumination \cite{Doherty2013a,Bourgeois2015a}. This conversion pathway highlights the close relationship between the spin, the charge state, and the shelving dynamics which has eluded a detailed description so far \cite{Siyushev2019a}.

This optical pathway connects the charge state dynamics with a shelving process in the long lived singlet state. In the (optical) high-power limit, this mechanism enables efficient charge state initialization via NV$^-$ to NV$^0$ transfer and subsequent electron recapture (Fig \ref{fig:steady_state}b). The model indicates that the negative charge state population can thereby be increased from $\sim80\,\%$ at low power to to over $92\,\%$ (see Appendix B). However, increased shelving under high-power illumination, i.e. far beyond saturation of the optical transition, would lead to continuously decreased luminescence with higher laser powers, while it has been shown that the NV center exhibits almost power-independent luminescence in this regime \cite{Han2012,Chapman2012}. This behaviour is accounted for in the model by allowing excitation out of the singlet state via single-photon absorption to a higher lying state (level 8 in Fig \ref{fig:steady_state}a) \cite{Bockstedte2018, Han2012}. Including this state, the luminescence is expected to slightly decrease past the saturation point while remaining almost constant at higher laser powers, consistent with the anomalous saturation behaviour observed previously \cite{Han2012, Chapman2012, Chen2015b}. Nonetheless, high power illumination overall provides a method to significantly improve the charge state purity.

After shelving, the NV$^-$ $m_s=0$ occupation will be limited to only $81\,\%$ of the total population, by a combination of the branching ratio of the decays from, and excitation out of the metastable state, according to our model (see Appendix B). Using high power initialization, we therefore sacrifice spin state purity to improve the charge state ratio. However, the $m_s=0$ population can now be increased by using weak excitation: For sufficiently low laser power, charge state conversion becomes negligible as it is suppressed due to its quadratic dependence on the illumination intensity \cite{Waldherr2014a} (Fig \ref{fig:steady_state}c). The spin polarization is then constrained by the ratio of the spin mixing rates, with a limiting value of $98\,\%$ in the $m_s=0$ state within the NV$^-$ manifold, resulting in $>90\,\%$ of the total population residing in the $m_s=0$ state. In comparison, conventional initialization results in only $\sim 77\,\%$ of the total population residing in $m_s=0$ (see Appendix B). Thus, using sufficiently weak laser pulses, two-power initialization is able to improve both charge and spin state purity. However, it should be noted that in practice the achievable spin polarization will be a compromise between expediency and charge purity.

\section{Experimental Results}\label{sec:results}


\begin{figure}[!t]
	\centering
	\includegraphics[width=\columnwidth]{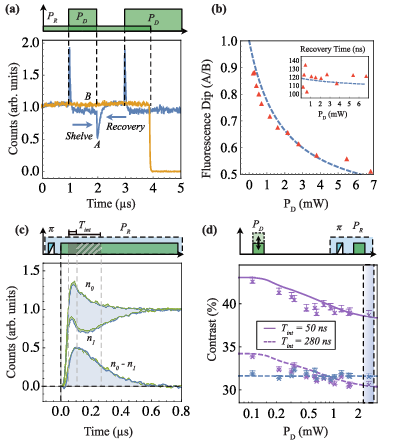}
	\caption{High-power dynamics. \textbf{(a)} Photon collection during a $P_R=0.6$ mW  readout pulse with (blue) and without (orange) additional $P_D=7$ mW pulses. The sudden decrease in fluorescence after switching off the strong pulse is indicative of shelving into a non-radiative state. \textbf{(b)} Depth of the fluorescence dip over $P_{D}$ (red dots) immediately after switching off the high-intensity laser. The dashed blue line is retrieved from our model which reproduces the increased dip with laser power well. At higher laser powers, the model curve predicts a minimum value of 0.35 for the fluorescence quench. The inset shows the recovery time extracted from an exponential fit to the luminescence traces for different $P_D$. The dashed line is the recovery time obtained from the model. \textbf{(c)} Comparison of spin contrast luminescence traces for $P_D=P_R=0.6\,$mW (blue) and $P_D=2\,$mW with $P_R=\SI{0.6}{\milli\watt}$ (green), where the latter corresponds to the highlighted data points in (d).The fluorescence counts collected with and without application of microwave $\pi$-pulse are denoted as $n_1$ and $n_0$ while the lower graph corresponds to the signal difference between the two states. \textbf{(d)} We now apply the high intensity laser pulse before the readout pulse and investigate the obtained contrast with increasing initialization power $P_D$ using a pulse length of \SI{3}{\micro\second}. Increasing $P_D$ while keeping the readout power constant results in a decrease of the contrast (purple). We additionally show the reference contrast (blue) obtained from conventional initialization using a \SI{0.6}{\milli\watt} initialization and readout pulse. We show both the maximal contrast (\SI{50}{\nano\second} integration window) and the contrast for the highest signal-to-noise ratio (\SI{280}{\nano\second}), from measurements (points) as well as numerical solutions from the model (lines). These integration windows are indicated by $T_{int}$ in the previous panel.}
	\label{fig:regular_trace}
\end{figure}

For our measurements, we use a home-built ODMR setup with two \SI{520}{\nano\meter}, \SI{80}{\milli\watt} laser diodes (Roithner LD-520-80MG). Two separate diodes are used here in order to guarantee that the readout pulses for different sequences are not affected by thermal changes, or any other memory effects in the diode which could be caused by different driving conditions. The diodes are driven by a directly TTL switchable diode driver designed for spike-free switching (IC Haus HG1D). One laser diode is used for readout pulses only, while the second is used for the high and low intensity laser pulses. Their beams are combined on a polarizing beamsplitter and coupled into a single-mode, polarization-maintaining fiber with orthogonal polarizations. However, we note that two-power initialization can be achieved with a single laser diode.

As they have orthogonal linear polarizations, the two lasers couple differently to the NV center: The readout laser excites the optical transition 1.4 times more efficiently than the initialization laser. We use a readout laser power of $P_R=\SI{0.6}{\milli\watt}$, typical for NV center experiments, and initialization laser powers $<\SI{21}{\milli\watt}\,$. Furthermore, in order to guarantee that our results are not altered by diode switching characteristics, we sample the incident beam and record time-resolved traces for both NV center fluorescence and laser power (see Appendix F).

The beam is focused on the sample by using an immersion-oil objective (Nikon CFI Plan Apo NCG 100X Oil) with a numerical aperture of 1.4 which is used for excitation as well as for photon collection. For our optical system, the nominal intensity at the NV center is $\sim \SI{20}{\milli\watt \micro\meter\tothe{-2}}$ per milliwatt at the input of the microscope objective. Our model indicates that the resulting excitation rates are \SI{35}{\mega\hertz\per\milli\watt} and \SI{25}{\mega\hertz\per\milli\watt} for the readout and initialization laser, respectively.

NV$^0$ fluorescence is partially filtered out from the collected light by using a 650 nm long-pass filter. The luminescence is collected with two optical fibers (Thorlabs P5-SMF28) to identify single NV centers, and detected by two avalanche photodiodes. Microwave pulses are produced with a signal generator (Analog Devices ADF4351), amplified (Minicircuits ZHL-16W-43+), and reach the NV center via a wire spanned across the sample.

We operate at an external magnetic field close to 0 G, where the $m_s = \pm 1$ states can be treated as degenerate, though we note that the scheme is equally applicable at larger fields, including those commonly used to polarize the nitrogen nuclear spin \cite{Jacques2009}. We further apply microwave pulses with a Rabi frequency of $2\pi \times \SI{13}{\mega\hertz}$, which is sufficient to efficiently drive all three transitions present due to hyperfine coupling between the nitrogen nucleus and the NV center's electronic spin \cite{Doherty2013a}. 

\subsection{Population Shelving and Contrast}

In order to investigate the predicted reduction in luminescence due to shelving after high power illumination, we use a $P_R=\SI{0.6}{\milli\watt}$ readout laser pulse supplemented by $P_D=\SI{7}{\milli\watt}$ pulses (Fig \ref{fig:regular_trace}a). After a brief spike in the count rate, part of the luminescence is quenched by the high-power excitation. Switching off the strong laser pulse leads to a further, rapid decrease, after which the luminescence recovers on a timescale dependent on the metastable state's lifetime and the power $P_R$. No such behaviour is observed when switching off the readout laser during the second strong pulse, corroborating the purported shelving due to intense illumination. Fig \ref{fig:regular_trace}b shows the power dependence of the shelving up to $P_D=\SI{7}{\milli\watt}\,$. At even higher laser powers than measured here, our model predicts a minimum value of 0.35 for the fluorescence quench.

We compare the fluorescence traces obtained with and without application of a microwave $\pi$-pulse (Fig \ref{fig:regular_trace}c), from which we directly calculate the respective readout contrasts. We define the contrast of an optically detected magnetic resonance (ODMR) signal as $C = 1 - S_{1}/S_0$, where $S_{1}$ and $S_{0}$ correspond to the total photon counts in the readout window from $m_s=\pm 1$ and $m_s=0$, respectively\footnote{This relates to the fringe contrast commonly used in e.g. optics, $C_{opt}=(S_{0}-S_{1})/(S_{0}+S_{1})$, through $C_{opt}=C/(2-C)$ \cite{Barry2020}.}. While a short integration time $T_{int}$ near the start of the readout gives the highest contrast (Fig \ref{fig:regular_trace}c), a longer readout duration will increase the collected photon number. However, prolonged illumination leads to charge state conversion as well as spin depolarization, and thereby to a reduction of the contrast over time. We thus provide two distinct values: maximal contrast is obtained from integrating over 50 ns near the start of the readout pulse. The contrast for a high signal-to-noise ratio (SNR) is instead obtained by integrating over a 280 ns window.

Fig \ref{fig:regular_trace}c, shows two spin state readout traces for two different initialization intensities, as well as the integration windows used to calculate the contrast values. The initialization powers used here are indicated by the blue area in Fig \ref{fig:regular_trace}d. The blue trace shows our result for conventional initialization, while green lines are obtained from traces using \SI{3}{\milli\watt} of initialization power. Here we obtain a slightly increased total luminescence signal, hinting at the increased charge state fidelity using high power initialization. Furthermore, the reduction in readout contrast with laser power is depicted in Fig \ref{fig:regular_trace}d, where we vary the initialization power $P_D$ while keeping the readout power fixed at $P_R=\SI{0.6}{\milli\watt}$. 

\subsection{Two-Power Initialization}

\begin{figure}[!t]
	\centering
	\includegraphics[width=\columnwidth]{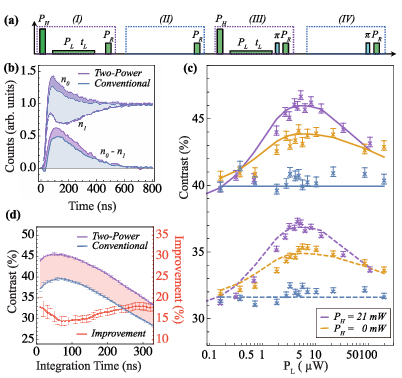}
	\caption{Two-power enhancement. \textbf{(a)} Sequence used during measurements consisting of 4 steps (I...IV), for direct comparison of two-power (I, III) and conventional (II, IV) initialization, where the system is read out using the same laser power as for initialization. \textbf{(b)} Counts over time during a readout pulse for conventional initialization (blue) and with two-power initialization (purple). We observe a clear increase in the obtainable contrast at the start of the readout pulse, which reduces over time due to increased spin mixing. \textbf{(c)} Sweep of the power in the low-power pulse with a duration of \SI{90}{\micro\second} after application of a $P_H=\SI{21}{\milli\watt}$ charge state pulse (purple) and when omitting the latter (orange). Furthermore we include the contrasts obtained from the conventional readout-pulse initialization for reference (blue). While dots indicate data points, lines indicate numerical solutions from the model. We show both maximal (solid lines) and best SNR (dashed lines) contrasts and for this duration obtain an optimal $P_L\approx \SI{6}{\micro\watt}$. \textbf{(d)} Resulting contrast over increasing integration window duration for our two-power (purple) and conventional (blue) initialization. The red line shows the improvement evaluated for each integration time which surpasses $17\,\%$ for integration windows longer than \SI{200}{\nano\second}.}
	\label{fig:low_power}
\end{figure}

We now initialize our system using a two-power scheme. The system is initialized into the negative charge state via trapping in the metastable state by application of a high-intensity laser pulse with a power  $P_H=\SI{21}{\milli\watt}$. We then apply a low-intensity pulse with power $P_L= \SI{6}{\micro\watt}$ and length $t_L=\SI{90}{\micro\second}$ to polarize the spin state via the more favorable excited-state branching ratios within NV$^-$, followed by a readout pulse (see (I) in Fig \ref{fig:low_power}a). We additionally perform measurements with the conventional (single-pulse single-power) readout pulse initialization method (see (II) in Fig \ref{fig:low_power}a).

\begin{figure}[!t]
	\centering
	\includegraphics[width=\columnwidth]{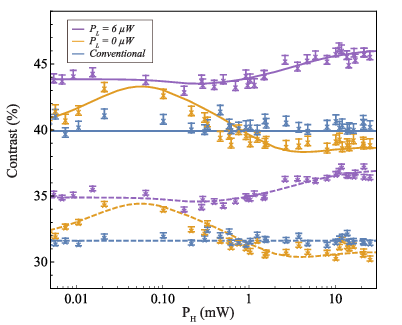}
	\caption{Contribution of the charge initialization pulse. High-intensity laser power sweep for $P_L=\SI{6}{\micro\watt}$ (purple) and $P_L=\SI{0}{\micro\watt}$ (orange) for maximal contrast (solid) and best SNR contrast (dashed). We additionally show our obtained contrast using conventional readout-pulse initialization (blue). Solid lines are obtained from solving the model while dots indicate data points.}
	\label{fig:two_pulse}
\end{figure}

In order to quantify the effects of the charge state initialization pulse (with power $P_H$), the spin initialization pulse (power $P_L$ and length $t_L$), we first fix $P_H=\SI{21}{\milli\watt}$ and $t_L=\SI{90}{\micro\second}$ while varying the power $P_L$ (see Fig \ref{fig:low_power}c). We observe a fairly wide optimum around $P_L=\SI{6}{\micro\watt}$ for which a maximal contrast around $46\,\%$ can be reached using the two-power sequence, compared to $\sim 43.5\,\%$ using only the low-power pulse.

Having chosen a power $P_L$, we aim to quantify the charge state initialization by varying the power of the high intensity laser, as depicted in Fig \ref{fig:two_pulse}. We perform these measurements once for $P_L=\SI{6}{\micro\watt}$ (purple) and once for $P_L=0$ (orange). Our results show a clear drop in contrast with increased $P_H$ for $P_L=0$. For data points using two-power initialization ($P_L = \SI{6}{\micro\watt}$), towards $P_H\rightarrow \SI{0}{\micro\watt}$, the low intensity pulse and our readout laser are responsible for initialization and tend towards $43.5\,\%$, similar to the results previously shown \cite{Song2020}. For initialization using solely the combination of readout pulse and high intensity pulse (orange), we observe a lower contrast optimum around $P_H=\SI{50}{\micro\watt}$ before dropping to the same contrast given by our reference measurements at $P_H = 0$.

We further perform ODMR measurements using the two optimal pulse parameters (Fig \ref{fig:high_speed}a). This was performed with and without two-power initialization, both implemented in the same sequence to guarantee identical external circumstances. It can be seen that the resonance minimum reaches a slightly lower value, while the off-resonant luminescence level is significantly higher, when using two-power initialization. We attribute the increase in photon counts to improved NV$^-$ population caused by the high-power pulse, while the lower minimum additionally underpins the high spin state purity obtained from the low-power pulse.

\section{Improvement for any sequence length}

\begin{figure}[!t]
	\centering
	\includegraphics[width=\columnwidth]{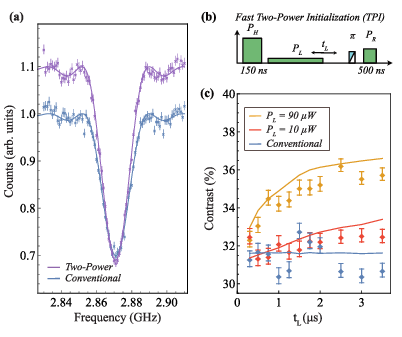}
	\caption{\textbf{(a)} ODMR measurement performed for two-power initialization using optimized parameters (purple) with $t_L=\SI{90}{\micro\second}$ and a readout laser power of \SI{0.9}{\milli\watt}. For comparison a sequence using conventional initialization where the system is read out using the same laser power as it is initialized with is shown in blue. Counts are obtained in \SI{280}{\nano\second} readout windows. Solid lines are sinc$^2$-fits to the data. \textbf{(b)} Depiction of the pulse sequence and pulse duration used for high-speed two-power initialization. \textbf{(c)} Contrast obtained from an optimized pulse sequence where the readout pulse is immediately followed by a short (150 ns) high intensity pulse $P_H=\SI{21}{\milli\watt}$ allowing for rapid charge state initialization. A \SI{200}{\nano\second} long pause is inserted for the laser diode to completely switch off before a low intensity pulse of varying power and length is applied. For a combined initialization and readout duration of \SI{3}{\micro\second} we already obtain a relative contrast increase of $12\,\%$ as compared to conventional initialization.
	}
	\label{fig:high_speed}
\end{figure}

We find from the model that implementing the two-power initialization sequence is always advantageous in terms of measurement sensitivity, independently of the free evolution time in the measurement sequence (see Appendix D). As an example, a speedup factor of 1.2 is predicted for a free evolution time of only \SI{500}{\nano\second}. Confirming this prediction, a measurement sequence with short initialization times is shown in Fig. \ref{fig:high_speed}b. For short evolution times, on the order of microseconds, relevant for recently developed methods such as QDyne detection \cite{schmitt2017submillihertz}, the speedup benefits chiefly from the fast charge state purification using high power pulses. We observe that the contrast is already improved, compared to conventional initialization, for the shortest measured sequence with an initialization duration of \SI{650}{\nano\second}. 
A far greater improvement can be achieved with only a modest increase of the initialization time: A maximal (SNR) contrast of $44\,\%$ ($35\,\%$) is already reached with a total combined initialization and readout time of only \SI{3}{\micro\second} with $P_L=\SI{90}{\micro\watt}$, $t_L=\SI{2}{\micro\second}$ and a readout power of $P_R=\SI{0.7}{\milli\watt}\,$ (Fig \ref{fig:high_speed}c). For comparison, conventional initialization using this readout power reaches $40.8\,\%$ ($31\,\%$) at the readout power used here. 

Longer measurement sequences, with a duration of tens to hundreds of microseconds, are regularly used in e.g. high-resolution magnetometry \cite{Barry2020}, the detection of external spins \cite{Lovchinsky2016,Arunkumar2021} using near-surface NV centers \cite{Sangtawesin2020}, or those involving the manipulation of nuclear spins \cite{rong2015}. Longer initialization times can be used in such sequences, which can thereby be accelerated with a speedup factor of $>1.5$ (see Appendix D).

\section{Summary}

In summary, we have shown that a two-power initialization sequence strongly increases the initial spin contrast of the NV center. The improvement persists over the course of the readout duration, giving a contrast increase of $>17\,\%$ after \SI{250}{\nano\second}. The benefit of the sequence is threefold: Firstly, the spin and charge states are initialized with high purity. Secondly, the luminescence collected in the NV$^-$ spectral window is increased, further improving the SNR. Thirdly, initialization times can be shortened down to only a couple of microseconds while still benefiting from the first two advantages. An additional, though here small, beneficial effect is the observed suppression of background luminescence by bleaching within the detection volume, the nature of which is yet to be determined (see Appendix E). 
Additionally we supply a detailed model that is in good agreement with the data recorded here and is able to predict the spin contrast, fluorescence behaviour and time evolution of the system under illumination.
The demonstrated method has two requirements: Following a standard readout pulse, a high-power laser pulse is required which must be switched off on a timescale much shorter than the metastable-state lifetime (i.e. $\ll \SI{200}{\nano\second}$). This pulse can be short (on the order of nanoseconds) since the driven shelving pathway does not involve any optical decay processes. Then, a low-power laser pulse with an intensity on the order of 0.5$\,\%$\textendash$10\,\%$ of the readout pulse is needed, the length of which is set by the desired improvement. Both requirements can be fulfilled in most standard NV center experimental setups using acousto-optic modulators or switched laser diodes \cite{Song2020,Oeckinghaus2014}.

\section{Discussion and Conclusion}

The two-power initialization method provides a direct route to improved measurements in almost any setting for nitrogen-vacancy sensors, and can be adapted for a large variety of spin centres \cite{Barry2020}. Additionally, the improved initialization may be of use even in low-temperature applications, where accelerated initialization routines for the spin and charge states can reduce the sequence duration \cite{Robledo2011,vasconcelos2020}.
The spin readout can be improved further by using a higher readout power than what was available in our experiment. Our model predicts a maximum contrast approaching $48\,\%$ using two-power initialization,  while a (hypothetical) perfect initialization procedure would produce a maximum contrast of slightly above $52\,\%$.
Finally, we note that the intricate interplay of the spin polarization with the charge state dynamics indicates that the mechanism implemented in this work will depend on the excitation wavelength \cite{Aslam2013, Storteboom2015}. This matter underlines the necessity for further investigation, by both theoretical modeling and experimental characterization of the NV center's response to excitation. 

\section{Acknowledgements} Financial support was provided by the FWF project I 3167-N27 SiC-EiC, FFG QuantERA 864036 Q-Magine, FFG 870002 QSense4Life, FFG 877615 QSense4Power, as well as the European Union’s Horizon 2020 and Horizon Europe research and innovation programmes under projects 101038045 (ChemiQS) and 101046911 (QuMicro). A.G. acknowledges support from the National Research, Development and Innovation Office in Hungary (NKFIH) Grant No. KKP129866 (National Excellence Program), and the EU QuantERA II MAESTRO project. A.G. and V.I. acknowledge support from the Quantum Information National Laboratory sponsored via the Ministry of Culture and Innovation of Hungary. V.I. acknowledges the support from the MTA Premium Postdoctoral Research Program and the Knut and Alice Wallenberg Foundation through WBSQD2 project (Grant No. 2018.0071) and the National Research, Development and Innovation Office in Hungary (NKFIH) Grant No. FK 137918. M.N. acknowledges the support from FWO (Funds for Scientific Research) Flanders, Projects no: G0D1721N and G0A0520N. M.G. acknowledges project No. 101038045 (ChemiQS): This project has received funding from the European Union's Horizon 2020 research and innovation programme.


\section{Appendix A: Model}

Having demonstrated the validity of two-power initialization, we aim to explore the effect of charge state conversion on the observed photo-dynamics of the defect, and infer its effect for higher laser powers. Building upon insights gained from previous work \cite{Aslam2013,BarGill2018,Song2020,Gaebel2006a}, we therefore construct an effective model for the dynamics (see Fig. \ref{fig:steady_state}).

We model the negative charge state as a five-level system including two ground states ($L_1$, $L_3$), two optically excited states ($L_2$, $L_4$) and one effective singlet state ($L_7$), driven by off-resonant green laser excitation. Levels $L_5$ and $L_6$ represent the neutral charge state (NV$^0$). Optically excited NV$^-$ states can either decay back to the NV$^-$ ground state via emission of a red photon, or decay to the singlet state ($L_7$) via a non-radiative and spin-dependent process. Excitation with green (\SI{520}{\nano\meter}) illumination does not supply sufficient energy to directly excite the system from the the triplet ground states to the conduction band. 
Charge state conversion occurs either from the optically excited levels in the triplet manifold of NV$^-$ \cite{Aslam2013} or by excitation from the metastable state ($L_7$). The driven channel from $L_2$ and $L_4$ takes place via either an Auger process \cite{Siyushev2013} or direct ionization \cite{Alkauskas2021} responsible for electron ionization. The former process decays into the ground state of NV$^0$ whereas the latter process scatters to the shelving $^4A_2$ state of NV$^0$ leaving one electron elevated to the conduction band. The direct photoexcitation can be approximated by an $er$-operator, which is a one-body operator. According to the Slater-Condon rules \cite{Slater1929, Condon1930}, only one spin orbital can change in this optical transition, therefore $^3E$ of NV$^-$ transforms to $^4A_2$ of NV$^0$ plus an electron in the conduction band by photoexcitation. In our experiments we do not find evidence of measurable population of the shelving $^4A_2$ state, therefore, it is ignored in the model. Conversely, the Auger-process can be described as swapping two spin orbitals, a two-body operator, therefore, $^3E$ of NV$^-$ can scatter to the ground state of NV$^0$ plus an electron high in the conduction band \cite{Siyushev2013}.

The photoexcitation of the $^2A_2$ excited state of NV$^0$ may also occur by either a direct process (promoting an electron from the valence band to the empty $a_1$ defect level in the gap) or an Auger-process (occupying the in-gap $a_1$ hole by an electron from the in-gap $e$ level and then promoting an electron from the valence band to the empty $e$ defect level in the gap). Both processes leave a hole in the valence band. In the direct process, the system arrives at the ground state of NV$^-$, i.e., $L_{6}$ goes to the $L_{1}$ and $L_{3}$ states because of the alluded Slater-Condon principle. On the other hand, the Auger-process enables to arrive at the metastable state $L_{7}$ of NV$^-$ too, besides the $L_{1}$ and $L_{3}$ states, where $L_7$ amalgamates the $^1A_1$ and $^1E$ singlet states because of the very short lifetime of the $^1A_1$ state \cite{Ulbricht2018}. The energy cost of these processes varies with the final state. The calculated adiabatic acceptor charge transition level of the NV defect is at about \SI{2.75}{\electronvolt} from the conduction band edge \cite{Deak2010, Londero2018}, whereas the calculated energy gap between the  ${}^3A_2$ ground state and ${}^1A_1$ state is at about \SI{1.6}{\electronvolt} (see Ref. \cite{Gali2019} and references therein). The total energy cost to convert the NV$^0$ ground state to the ${}^1A_1$ NV$^-$ excited state is then about \SI{4.3}{\electronvolt} which coincides with twice the zero-phonon-line (ZPL) energy of NV$^0$. This means that special excited state of $^1A_1$ of NV$^-$ binding a hole resonant with the valence band maximum develops. This hole is Coulombically bound, which is a special bound exciton state or Rydberg state which has been observed for the SiV defect \cite{Zhang2020} and has been recently implied and modeled for the ${}^3A_2$ plus a bound hole system for the NV$^-$ defect \cite{Mizuochi2012, Meriles2021}. The bound hole is loosely localized following the effective mass theory. By even taking into account the possible relaxation energy of the ions caused by the change in the electronic states, we may claim that \SI{520}{\nano\meter} laser excitation can reach the $^1A_1$ plus bound hole state of NV$^-$ by two-photon excitation of NV$^0$. Scattering to the $^1E$ ($L_7$) and $^3A_2$ ($L_1$ and $L_3$) states of NV$^-$ via an Auger-process leaves a hole deep in the valence band at around \SI{1.2}{\electronvolt} and \SI{1.6}{\electronvolt} from the valence band maximum, respectively. According to our calculation, a resonant $a_1$ state, broadened by the diamond bands occurs in this energy region which originates from the dangling bond orbitals of the carbon and nitrogen atoms near the vacant site. Unlike the usual diamond bands that are completely delocalized, the resonant state is weakly localized. This should lead to a larger direct and Auger-ionization rate of NV$^0$ than those of NV$^-$, because no such high-energy resonant state exists in the conduction band, critical in the photoionization of NV$^-$. Previous calculations on the ionization rates of NV$^-$ implied that the Auger-rates are significantly faster than the direct ionization rates \cite{Siyushev2013}. By asserting the same scenario for the photoionization of NV$^0$ and considering the resonance condition towards the $^1A_1$ state ($L_7$), we assume a decay from $L_{6}$ to $L_{7}$ and existing decay channels towards $L_{1}$ and $L_{3}$. In this case, the NV$^-$ ground state triplet is spin polarized towards the $m_s=0$ state.
Nevertheless, simulation results could not entirely exclude that direct ionization to the ground state occurs. In this case, the group theory analysis imply that the relative population of the $m_s=0$:$m_s=\pm1$ states of $^3A_2$ is given by 1/3:2/3.

Furthermore, at high excitation powers, the NV center exhibits nearly power-independent luminescence \cite{Chapman2012, Han2012}. This behavior is incompatible with perfectly "dark" shelving of the electron in the metastable state, since such a system would display decreasing luminescence towards higher excitation intensity. We therefore include excitation from the metastable ground state ($L_7$) via a single-photon transition to a higher lying state ($L_8$). A strong and broad transition around \SI{2.58}{\electronvolt} from $^1E$ to $^1E'$ has been observed in numerical simulations and could be excited by \SI{520}{\nano\meter} green illumination (\SI{2.384}{\electronvolt}), providing a possible candidate for such a mechanism \cite{Bockstedte2018}. $^1E'$ can decay to the triplet excited states via an inter-system crossing \cite{Maze2011,Gali2019}. In the model, decay from $L_8$ is allowed towards the NV$^0$ ground state and to the NV$^-$ triplet excited states, but the numerical optimization converges towards values with negligible decay to the neutral state. We underline that the excitation efficiencies for many of these processes are expected to be strongly wavelength-dependent, leading to different dynamics for other excitation energies \cite{Storteboom2015}.

The processes described above are collected in a rate equation system where $\sigma_{i,j}$ denotes the excitation cross section for a transition $L_i \rightarrow L_j$, scaled via a variable $P$ which is globally varied by two free parameters determining laser coupling to the NV center. Since lasers with orthogonal polarization can couple differently to the NV center, we adjust the driving rate constant $P$ independently for each laser.
Similarly, the decay rate from an excited state $L_e \rightarrow L_g$ is given by $\Gamma_{e,g}$. We fix a number of decay rates to literature values, namely $\Gamma_{2,1}$, $\Gamma_{4,3}$ and $\Gamma_{6,5}$ (see Table \ref{tab:model_parameters}). Furthermore, we fix the lifetime of the higher lying state ($L_8$) to \SI{200}{\pico\second} and only vary it's relative branching ratios into $L_1$, $L_3$ and $L_5$. Since not much information about the details of this Rydberg state is known, we opted to assign a comparatively fast decay rate to it. This ensures that no population is shelved in this state while it's branching still remains included. Our fitted model suggest that this Rydberg state decays equally into to the optically excited $m_s=0$ and $m_s=\pm 1$ states, however, the obtained error bars leave the possibility of potential spin polarization open. Here, further investigation might be of interest and lead to a better understanding of the optical excitation cycle of the NV center.

Additionally, we fix the excitation cross sections $\sigma_{3 ,4} = \sigma_{1,2} = 1$ while all other excitation cross sections are extracted from a numerically optimized fit (see Appendix G) to the data shown in the main text, and are listed in Table \ref{tab:model_parameters}.

\begin{table}[t]
	\renewcommand{\arraystretch}{1.2}
	\begin{center}
		\begin{tabular*}{\linewidth}{@{\extracolsep{\fill}} l c}
			\hline \hline
			Parameter & Value \\
			\hline
			\multicolumn{2}{c}{Decay time (ns)}\\
			$\Gamma^{-1}_{2,1}$ & 13 \\
			$\Gamma^{-1}_{2,7}$& $93.5(8)^{*}$\\
			$\Gamma^{-1}_{4,3}$& 13 \\
			$\Gamma^{-1}_{4,7}$& $14.98(6)^{*}$ \\
			$\Gamma^{-1}_{6,5}$& $20$ \\
			$\Gamma^{-1}_{7,1}$& $186.12(24)^{*}$ \\
			$\Gamma^{-1}_{7,3}$& $2722(57)^{*}$ \\
			\multicolumn{2}{c}{Normalized cross section}\\
			$\sigma_{1,2}\equiv\sigma _{3,4}$& 1 \\
			$\sigma_{2,5}\equiv\sigma _{4,5}$& $0.237(4)^{*}$\\
			$\sigma_{7,8}$ & $0.0059(3)^{*}$ \\
			$\sigma_{5,6}$ & $0.562(7)^{*}$\\
			$\sigma_{6,7}$ & $0.15(1)^{*}$\\
			$\sigma_{6,1}$ & $0.281(7)^{*}$ \\
			$\sigma_{6,3}$ & $0.3(2)^{*}$ \\
			\multicolumn{2}{c}{$\mathrm{L_8}$ branching}\\
			$\Gamma_{8,4} / \Gamma_{8,2}$ & $2.7(1.6)^{*}$ \\
			$\Gamma_{8,5} / \Gamma_{8,2}$ & $0^{*}$\\
			\multicolumn{2}{c}{Power scaling (MHz/mW)}\\
			Readout Laser & $35.0(3)^{*}$  \\
			Initialization Laser & $25.0(2)^{*}$\\
			\hline \hline
		\end{tabular*}
	\end{center}
	\caption{Model parameters used for the rate equation model. Fitted parameters obtained from the results in the main text are marked with an asterisk.}
	\label{tab:model_parameters}
\end{table}

\section{Appendix B: Model predictions}\label{subsec:predictions}

In this section, we evaluate our model in the high- and low-power excitation regimes, highlighting several features that are of relevance for the measurement of spin-dependent luminescence and photo current. The system presents power-dependent shelving due to $\sigma_{6,7}$.  At low powers, the metastable state will be depopulated predominantly via decay into the triplet ground states ($\Gamma_{7,1}$ and $\Gamma_{7,3}$). At higher powers, the shelved population is reduced by laser excitation via the transitions with cross section $\sigma_{7,8}$. The balance between these processes will define the anomalous saturation, since conversion from NV$^-$ to NV$^0$ results in an increase of the shelving rate, while excitation out of the metastable state leads to fluorescence in the high-power limit. These rates also define the photoelectric signal under green illumination, and negate the possibility of perfectly protecting nuclear spins from electron decoherence using high-power laser illumination \cite{Maurer2012}.

These mechanisms also explain our improved charge state initialization: under sufficiently high driving powers, population will be strongly shelved in $L_7$, so long as the excitation rate of the singlet state remains sufficiently small. We find that more than 80$\,\%$ of the population is pumped into the metastable state, such that the charge state error is reduced from $19.5\,\%$ to $8\,\%$. After subsequent relaxation, the shelved population will branch into the $m_s=\pm1$ state ($L_3$) with $R_{7,3}=\Gamma_{7,3}/(\Gamma_{7,1}+\Gamma_{7,3})=0.07$, and into the $m_s=0$ state ($L_1$) with  $R_{7,1}=\Gamma_{7,1}/(\Gamma_{7,1}+\Gamma_{7,3})=0.93$ (see Table \ref{tab:model_parameters}). A high-intensity pulse thus initializes the charge state effectively, at the cost of spin polarization within the NV$^-$ manifold.

\begin{figure}
	\centering
	\includegraphics[width=\columnwidth]{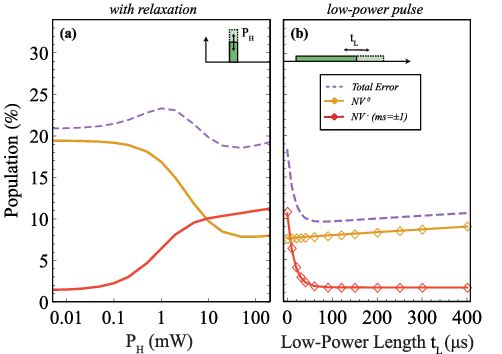}
	\caption{\textbf{(a)} Population after relaxation from the steady state for continuous-wave illumination with intensity $P_H$. High-power illumination leads to shelving in the metastable state, which permits high-fidelity charge-state initialization. The NV$^0$ population (orange) decreases towards high pumping power $P_H\gg1\,$mW, but the $m_s= \pm 1$ population (red) increases. The sum of these populations gives the total initialization error (dashed). \textbf{(b)} Model evaluation of spin state error within NV$^-$ (portion of total population in $m_s=\pm 1$), charge state error (population in NV$^0$) and total error (sum of both, purple) under low power initialization after a \SI{21}{\milli\watt} initialization pulse. The dots are obtained from low power approximations (see main text) which coincide well with the full model at low excitation powers.
	}\label{fig:init_error}
\end{figure}

At this point, the $m_s=0$ population can, however, be increased using weak laser excitation: In the low-power limit, where $P \sigma_{2,5}\ll \Gamma_{2,1}+\Gamma_{2,7}$ and $P \sigma_{4,5}\ll \Gamma_{4,3}+\Gamma_{4,7}$, the rate of transfers between charge states is negligible. It is therefore possible to initialize the spin state effectively within the NV$^-$ subsystem in this regime, via the decay rate ratios $\Gamma_{4,7}/\Gamma_{2,7}>1$ and $\Gamma_{7,1}/\Gamma_{7,3}>1$. The $m_s=0$ population can then be increased to a maximum of $1/\left(1+R_{2,7}\Gamma_{7,3}/R_{4,7}\Gamma_{7,1}\right)=98\,\%$ within the NV$^-$ manifold, where $R_{2,7}=\Gamma_{2,7}/(\Gamma_{2,1}+\Gamma_{2,7})$ and $R_{4,7}=\Gamma_{4,7}/(\Gamma_{4,3}+\Gamma_{4,7})$, respectively. Since time is a limiting factor in measurements, in practice the achievable spin polarization will be the result of a compromise between expediency and charge state conversion probability.

The key mechanisms are summarized in Fig.\ref{fig:init_error}. The left plot (Fig \ref{fig:init_error}a) shows the calculated populations for NV$^0$ (orange) and $m_s = \pm 1$ (red) for relaxation after a long pulse of a given power, i.e. after relaxation from the steady state. The dashed purple line is the total initialization error calculated by taking the sum of both. While in the low power regime we obtain a constant NV$^0$ population of about 19.5 \%, in agreement with the findings in \cite{Aslam2013}, increasing the power beyond \SI{1}{\milli\watt} will initialize the charge state towards NV$^-$ while simultaneously worsening the NV$^-$ spin state initialization. The right plot (b) shows the effect of subsequent spin initialization using low intensity versus time (solid lines). It can be seen that this second pulse rapidly reduces the spin error caused by the charge state pulse. For long illumination times, the error increases again due to slow charge state conversion of the NV center. We find that for low powers this rate can be approximated as $P^2\sigma_{\text{1,2}}\sigma_{\text{2,5}}/\left(\Gamma_{2,1}+\Gamma_{2,7}\right)$, and the polarization rate as $P \sigma_{1,2}R_{4,7}R_{7,1}$. Conversely, the NV$^0$ to NV$^-$ charge recapture can be approximated as $P^2 \sigma_{\text{5,6}}\sigma_{\text{6,7}}/\Gamma_{\text{6,5}}$. The results obtained from these approximations are indicated by the diamonds in Fig \ref{fig:init_error}b.

\begin{table*}
	\renewcommand{\arraystretch}{1.4}
	\centering
	\begin{tabular*}{\linewidth}{@{\extracolsep{\fill}} l c c c c c c c}
		\multicolumn{1}{c}{ } & \multicolumn{2}{c}{Population} & \multicolumn{3}{c}{Branching Ratio} & \multicolumn{2}{c}{Contrast}  \\
		\text{Parameters} & $L_0\text{(max)}$ & $L_1\text{(max)}$ & MS($R_{7,1}$) & E0($R_{2,7}$) & E$\pm$1($R_{4,7}$) & $L_0\text{=1}$ & $L_0\text{(max)}$ \\
		\hline \hline
		\text{Tetienne \cite{Tetienne2012}} & 0.88 & 0.12 & 0.57 & 0.09 & 0.41 & 0.38 & 0.3\\
		\text{Gupta \cite{Gupta16}} & 0.91 & 0.09 & 0.7 & 0.13 & 0.57& 0.51 & 0.44\\
		\text{Kalb \cite{Kalb2018}} & 1 & 0 & 0.8 & 0 & 0.41& 0.39 & 0.39\\
		\text{Thiering and Gali \cite{thiering2019eg}} & N.A. & N.A. & 0.84 & N.A. & N.A.& N.A. & N.A.\\
		\text{\emph{Chosen}} & 0.98 (0.79) & 0.017 (0.014) & 0.93 & 0.12 & 0.46 & 0.39 & 0.33\\
		\hline
	\end{tabular*}
	\caption{Comparison of selected and literature parameters. The first double-column shows the maximal $m_s=0$ and $m_s=\pm 1$ populations ($L_0$, $L_1$) in the low-power limit. Since selected literature models do not include both charge states of the NV center we state two values for respective $L_0(max)$ and $L_1(max)$ populations. The first value assumes no charge mixing, for which comparatively high spin initialization is obtained. Values in parentheses, instead, give the resulting population taking the neutral charge state into account. The second double column shows the metastable state branching ratio (MS) towards the optical ground state as well as the branching ratios of the optically excited NV$^-$ states. Furthermore using these values we calculate the maximal initial contrast ($L_0=1$) for which we assume perfect $m_s=0$ initialization as well as the spin readout contrast obtained from the maximally achievable $m_s=0$ initialization. We note that these values are lower than what we obtained in our measurements since, in order to allow for comparison between literature values, they are calculated as $C=1-\frac{L_0 R_{4,3} + L_1 R_{2,1} + 0.6 L_5}{L_1 R_{4,3} + L_0 R_{2,1} + 0.6 L_5}$, where the factor 0.6 stems from the spectral selectivity of NV$^0$.}
	\label{tab:literature_parameters}
\end{table*}

The resulting luminescence contrast and the spin polarization are both highly dependent on the parameters assumed for the branching ratios in the excited and metastable states. In Table \ref{tab:literature_parameters}, we compare selected values from the literature, cautioning that the assumptions, measurement methods, and experimental settings were rather different in each case. For each set, we also show the calculated initial contrast achievable for perfect spin and charge state initialization, as well as only for perfect charge state initialization.

\begin{figure}
	\centering
	\includegraphics[width=\columnwidth]{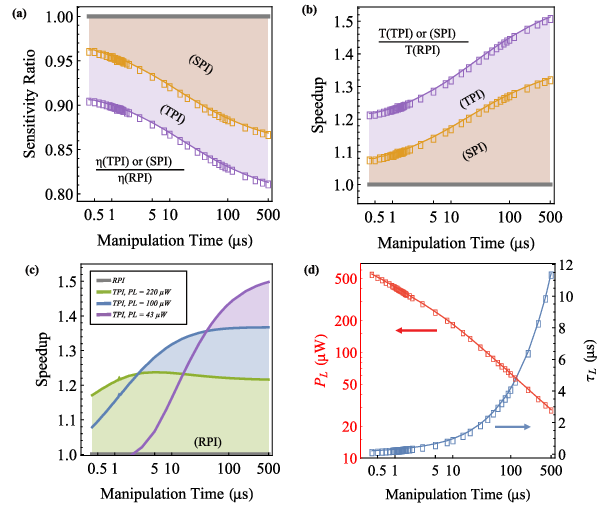}
	\caption{Readout power. Model contrast predictions under variation of readout power for three cases. Green lines show the result if the system could be perfectly initialized into $m_s=0$. Purple lines correspond to a charge state pulse of 20 mW, i.e. the value with the best expected performance in our data, followed by a low intensity pulse of \SI{6}{\micro\watt} for \SI{90}{\micro\second}. Orange lines show the conventional sequence used in experiments, where the readout pulse is of the same power as the initialization pulse. Solid lines show the maximal contrast obtained at a singular point within the readout duration while the dashed lines show the SNR contrast (for maximum SNR), found by optimizing the readout duration (red line).}
	\label{fig:contrast}
\end{figure}

\section{Appendix C: Luminescence Signal}\label{subsubsec:luminescence}
The maximum contrast and the contrast at the best signal-to-noise ratio (SNR) for spin-dependent luminescence measurements can be extracted from the time-dependent state occupations under pulsed excitation, and are shown in Fig. \ref{fig:contrast}. Firstly, we evaluate the dynamics of the system under the assumption that the populations have reached a steady state after illumination with laser intensity $P_R$ (scenario (I) in Fig. \ref{fig:contrast} )). This equilibrium is usually reached even after a few short measurement pulses on the order of $\sim$ \textmu s, and should therefore correctly depict common experimental conditions. Further, it is necessary to consider the relative signal strengths for the luminescence of the two charge states, which cannot be perfectly discriminated by spectral filtering. In the experiment, we use a long-pass filter with an edge wavelength of \SI{650}{\nano\meter}. From the filter transmission and the spectral properties of the two charge states \cite{Mizuochi2012}, we estimate a spectral selectivity of NV$^0$:NV$^-$ of 0.6:1. The combined luminescence signal of the two charge states is thus calculated from the bright transition rates $S=L_2\Gamma_{2,1}+L_4\Gamma_{4,3}+0.6\times L_6\Gamma_{6,5}$. 

In order to extract the contrast, we use two input states: in the first case, the population distributions after initialization are taken as the initial conditions, resulting in the signal $S_0$. In the second case, the $L_1$ population is fully swapped with the $L_3$ population, corresponding to a $\pi$-rotation of the initialized state from $m_s=0$ to $m_s= \pm 1$, resulting in the signal $S_1$. The optimal measurement time is then found by numerical integration of the two signals over the readout time to maximize the quantity SNR$=(S_0-S_1)/\sqrt{S_0+S_1}$.

We examine two further initialization conditions. In scenario (II), the starting conditions are chosen to match the population resulting from the two-step initialization sequence described in the main text with $P=\SI{20}{\milli\watt}$ and $t_L=\SI{90}{\micro\second}$. This choice improves both the maximal contrast and the best-SNR contrast significantly, reaching values of $48\,\%$  and $38\,\%$, respectively. Green lines show the upper contrast limit predicted by our model in scenario (III), where perfect initialization into $m_s=0$ is assumed. We make the simplifying assumption that the excitation light causes no extraneous background scattering. 

In summary, the two-step initialization procedure, combined with an optimized readout intensity, provides a remarkable improvement in the signal difference that can be extracted from a spin-dependent measurement. Our model furthermore predicts that, with further improvements to the initialization, a peak contrast of almost $52\,\%$ (best SNR: $41\,\%$) could be reached by perfect spin and charge state initialization.

\section{Appendix D: Efficiency}\label{sec:efficiency}

Two-power initialization can effectively increase the readout contrast by more than $\sim\,17\,\%$ and boost the peak photon count rate by $11\,\%$ (see main text). We will follow up with an analysis of the expected speed ups gained from using three separate laser powers for readout and initialization and compare our results to the conventionally used method of using a single pulse for readout and initialization. High power illumination will lead to an initial reduction in readout contrast as compared to the commonly used method due to the inferior branching ratio of the metastable state (see previous sections and main text). However, for the high intensity pulse $P_H$, only very short times are necessary since, according to our model, it only relies on driven channels which are not delayed by any decay time. Here we expect a trade off between expediency and initialization quality to occur. Conversely, the low intensity pulse $P_L$ must be sufficiently weak in order to avoid charge state conversion. This requirement, in turn, can lead to long (many \textmu s) pulses for the spin polarization to take effect. In order to obtain a break-even point that optimizes the applied initialization for a given manipulation time $\tau_M$ we compare three different initialization methods with respect to their sensitivities. We follow the treatment of \cite{Barry2020}, where the sensitivity is given by
\begin{equation}
\eta \sim \frac{\sqrt{\tau_I + \tau_M}}{\sqrt{N}} \frac{1}{C_{opt}} \frac{1}{\tau_M} \frac{1}{e^{-\tau_M/T^*_2}}.
\label{eq:sensitivity}
\end{equation}
Here, $T^*_2$ corresponds to the dephasing time, $\tau_M$ and $\tau_I$ represent the time necessary for manipulation and initialization, respectively, $C_{opt}=(S_1-S_0)/(S_1+S_0)$ is the readout contrast, and $N$ is the average number of photons collected per sequence. 

We numerically solve our model for three cases: readout pulse initialization (RPI) is the commonly employed method, where a laser pulse with power $P_R$ and duration $\tau_R$  provides both readout and initialization. Single pulse initialization (SPI) includes a second, weaker pulse of length and power $P_L,\,\tau_L$ for initialization, thus different powers are allowed for readout and initialization. For two-power initialization (TPI), we include a short, high-intensity pulse ($P_H,\,\tau_H$). We furthermore include a pause $\tau_P$ after each initialization sequence, so that metastable state decay into the NV$^-$ optical ground state can take place. Additionally, we define the system as "initialized" if, independently of the starting state, the system always ends up up in the same final state (up to a $0.1\,\%$ error). The total initialization time is then calculated as $\tau_I=\tau_R+\tau_H+\tau_L+\tau_P$ and optimized for all manipulation lengths, $\tau_M$. Since we compare the three cases for identical spin manipulation times, the factor $\tau_M \exp{(\tau_M/T^*_2)}$ drops out in the calculation.

To obtain the speedup provided by our improved initialization sequence, we compare readout-pulse initialization to single- and two-power sequences using
\begin{equation}
\frac{T_{(SPI)\text{ or }(TPI)}}{T_{RPI}}=\frac{\eta^2_{(SPI)\text{ or }(TPI)}}{\eta^2_{RPI}},
\end{equation}
where $\eta$ corresponds to the sensitivity of the respective initialization method at a fixed manipulation time $\tau_M$ and $T$ is calculated as $T=\eta^2$ and represents the time necessary to reach a certain SNR.

\begin{figure}
	\centering
	\includegraphics[width=\columnwidth]{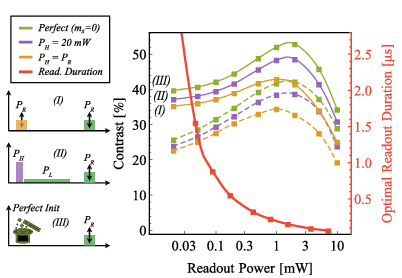}
	\caption{Speedup. \textbf{(a)} We optimize the sensitivity for RPI, SPI and TPI for a set of manipulation times to obtain the sensitivity gain. All parameters are left free during optimization though we force the application of a 20 ns long high power (30 mW) pulse for TPI. Both SPI and TPI prove beneficial for all manipulation lengths. \textbf{(b)} Predicted speedup in measurement time. \textbf{(c)} Comparison of the sensitivity for four selected cases: (gray) readout pulse initialization with $P_R=1\,$mW and two-power initialization (TPI) for $P_R=\SI{1}{\milli\watt}$ and fixed $P_L=\SI{380}{\micro\watt}$ (green), $P_L=\SI{200}{\micro\watt}$ (blue) and $P_L=\SI{43}{\micro\watt}$ (purple). With these fixed values we only allow $\tau_L$ to vary. \textbf{(d)} Optimal $P_L$ and $\tau_L$ for all sets of manipulation times using TPI (we note that SPI leads to very similar results).}
	\label{fig:speedup}
\end{figure}

The results are depicted in Fig \ref{fig:speedup} showing the scaled sensitivities for RPI and three cases of TPI. We obtain an immediate improvement in sensitivity for both SPI and TPI. The benefit from SPI shows that initialization and readout should in general be performed at different powers due to power dependent charge state conversion. The benefit of TPI stems from the fast state initialization at high powers. While a single readout pulse typically needs about 1-2 \textmu s to properly initialize the system, usually only the first 300 ns are of interest for signal acquisition. Using TPI on the other hand, the readout pulse can be kept short (300 ns) while subsequent application of the high power pulse only needs about 20 ns to reach a steady state. The spin state error resulting from high power initialization (see main text) is, however, outweighed by the comparatively fast initialization. For long manipulation times an increasing amount of time can be spend to repair the resulting spin error using subsequent low power illumination without significant increase in sequence length. Sensitivitiy for TPI thus gets monotonically better since the low power pulse can be kept at lower and lower powers while increasing its length to avoid charge state mixing. Fig \ref{fig:speedup}b shows the predicted speedup for TPI and SPI.
Furthermore, Fig \ref{fig:speedup}c shows the achieved speedup for three selected low power pulses. While the previous results were obtained by optimization of pulse lengths and powers, here we fix the readout power $P_R=\SI{1}{\milli\watt}$, and $P_L=\{\SI{43}{\micro\watt}, \SI{200}{\micro\watt}, \SI{380}{\micro\watt}\}$ while we numerically optimize the low power pulse length at these instances (Fig \ref{fig:speedup}c). For short manipulation times, it proves beneficial to apply short, comparatively strong pulses (380$\,\mu$W) at the end of the initialization step instead of long e.g. \SI{43}{\micro\watt} pulses since low power pulses require more time for proper initialization. For long manipulation times, however, we obtain a speedup surpassing 1.5. Fig \ref{fig:speedup}d shows the numerically calculated optimal powers and pulse lengths for the low power pulse in TPI. We note that optimal $P_L$ and $\tau_L$ for SPI derived from our model are almost identical to TPI.


\section{Appendix E: Background Subtraction}\label{sec:background}
\begin{figure}
	\centering
	\includegraphics[width=\columnwidth]{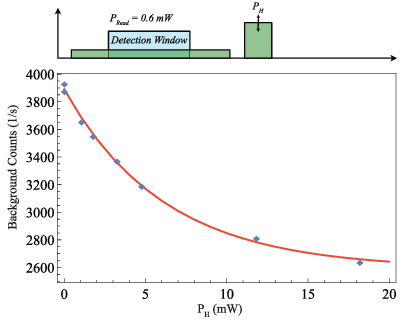}
	\caption{Background counts taken \SI{3}{\micro\meter} away from the NV center during a constant 0.6 mW readout pulse when followed by an additional \SI{5}{\micro\second} long pulse with varying power $P_H$. During these measurements, as in all data captured in the main text, we kept the total sequence length at \SI{100}{\micro\second} in order to maintain identical circumstances. We obtain an exponential reduction from $\sim\,3900\,$ counts per second to $\sim\,2700$ counts per second with increasing $P_H$. For comparison, at 0.6 mW we collect about $7\times10^4$ counts per second when focused on the NV center. We attribute the reduced fluorescence to a background signal that is bleached away by increasing illumination intensity.}
	\label{fig:background_subtraction}
\end{figure}
During all measurements, we observed a slight increase of 2$\,\%$ in the reference contrast from windows II and IV (see main text) when the high intensity pulse was added to the TPI sequence (windows I and III). We thus performed a set of background measurements using the two-power initialization sequence while varying the power of the high intensity pulse (Fig \ref{fig:background_subtraction}). These measurements were performed by collecting fluorescence \SI{3}{\micro\meter} away from the NV center on a spot without any other NV centers at the same depth.
We observed a clear decrease in collected background photons with increasing $P_H$, which we attribute to unwanted fluorescence that is bleached away by the additional pulse. We thus corrected our photon count rates using $P_H$-dependent background counts.

\section{Appendix F: Laser Characteristics}
\begin{figure}[ht]
	\centering
	\includegraphics[width = \columnwidth]{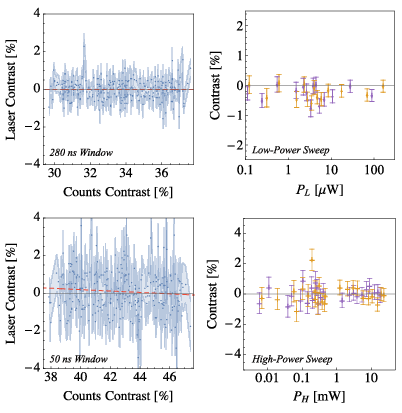}
	\caption{Laser contrast correlations. We perform a contrast calculation on the laser power traces recorded using a separate photon counter. The data processing is identical to that used for the evaluation of NV center photoluminescence counts. The left plots show the contrast that could stem from laser diode fluctuations versus the contrast from each data point for both long (top) and short (bottom) integration windows. Red lines in the plot show a linear fit through all points. Right: Contrast of our data sets that sweep $P_H$ and $P_L$ when applied to the laser traces.}
	\label{fig:laser_correlation}
\end{figure}
As described in the main text, we use two laser diodes driven directly by a TTL-switchable diode driver for readout and high/low intensity pulses, respectively. This separation prevents potential correlations caused by laser diode characteristics. As an example, driving a diode at peak power can cause intensity variations in subsequent pulses due to electronic or thermal effects.  We stress that the use of two laser diodes is not a requirement for the improved initialization, and was only implemented to guarantee highly repeatable readout pulses for a broad variation of initialization pulse characteristics. 

In order to quantify the effects of laser fluctuations on our results, we sampled part of the excitation light into a fiber to record the laser time traces during the experiments described in the main text. We applied all contrast calculations to our fluorescence traces (main text) as well as to the recorded laser traces. Fig \ref{fig:laser_correlation} shows the obtained contrast from the laser versus the contrasts resulting from our fluorescence traces (left) for both short and long integration windows. While for short integration windows we see a very slight anti-correlation (red lines), we observe no correlation between the two quantities for longer integration windows. We additionally apply our contrast calculation to the $P_H$ and $P_L$ sweeps. These also show no significant correlation, proving that laser pulse variability does not contribute to the observed contrast improvements.

\section{Appendix G: Numerical Optimization and Genetic Algorithm}
In order to fit our free parameters to the experimental results we numerically solve the system matrix to retrieve time traces for each data measurement trace. Those traces consist of the first 500 ns of each readout pulse for both readout power initialization, as well as two power initialization. Our optimization parameter is obtained by finding the least-squares deviation of the simulated fluorescence to the measured fluorescence. Additionally, we compare the fluorescence of our model during stepwise increase of continuous laser excitation power to a measured fluorescence. The sum of both is calculated to obtain the total fit error $\Delta_{Opt}$.
We further give an abstract overview of the optimization process: We first create a parent state vector $P_1=(A_1,A_2,...,A_n)=(\Gamma_1,...,\Gamma_j,\sigma_1,...,\sigma_k)$ for which we can evaluate $\Delta_{Opt}$. We additionally create a second parent $P_2$ that contains randomized elements. Both parents are used to create three offspring members $O_{1,2,3}$ via random cutting and swapping. Below we describe the procedure using a 7-parameter space. Offspring creation then leads to
\begin{equation}
\begin{multlined}
P_1 = \begin{bmatrix}
A_{1} \\ A_{2} \\ A_{3} \\ A_{4} \\ A_{5} \\ A_{6} \\ A_{7}
\end{bmatrix},
P_2 = \begin{bmatrix}
B_{1} \\ B_{2} \\ B_{3} \\ B_{4} \\ B_{5} \\ B_{6} \\ B_{7}
\end{bmatrix}
\\
\rightarrow
O_1 = \begin{bmatrix}
B_{1} \\ B_{2} \\ A_{3} \\ A_{4} \\ A_{5} \\ A_{6} \\ A_{7}
\end{bmatrix},
O_2 = \begin{bmatrix}
A_{1} \\ A_{2} \\ A_{3} \\ A_{4} \\ A_{5} \\ A_{6} \\ B_{7}
\end{bmatrix},
O_3 = \begin{bmatrix}
B_{1} \\ B_{2} \\ B_{3} \\ B_{4} \\ A_{5} \\ A_{6} \\ A_{7}
\end{bmatrix}
\end{multlined}
\end{equation}
Offspring vectors are then mutated such that each parameter has a probability to change by a random factor $\delta_i \in [1-\epsilon,1+\epsilon]$, e.g. $\Tilde{A}_3=\delta_3 \, A_3$. We give all elements of each offspring member a 20 $\%$ chance of mutation. From this procedure we obtain e.g. the set
\begin{equation}
\begin{multlined}
P_1 =
\begin{bmatrix}
A_{1} \\ A_{2} \\ A_{3} \\ A_{4} \\ A_{5} \\ A_{6} \\ A_{7}
\end{bmatrix},
P_2 =
\begin{bmatrix}
B_{1} \\ B_{2} \\ B_{3} \\ B_{4} \\ B_{5} \\ B_{6} \\ B_{7}
\end{bmatrix},
\\
O_1 =
\begin{bmatrix}
\Tilde{B}_{1} \\ B_{2} \\ \Tilde{A}_{3} \\ \Tilde{A}_{4} \\ A_{5} \\ A_{6} \\ A_{7}
\end{bmatrix},
O_2 =
\begin{bmatrix}
\Tilde{A}_{1} \\ \Tilde{A}_{2} \\ A_{3} \\ \Tilde{A}_{4} \\ A_{5} \\ A_{6} \\ B_{7}
\end{bmatrix},
O_3 =
\begin{bmatrix}
B_{1} \\ \Tilde{B}_{2} \\ B_{3} \\ B_{4} \\ A_{5} \\ A_{6} \\ \Tilde{A}_{7}
\end{bmatrix}
\end{multlined}
\end{equation}
for which we evaluate all individual $\Delta_{Opt}$, then select the two that minimize $\Delta_{Opt}$ best, which are chosen as new parents from which the process repeats. Finally we note that in the actual optimization process, six offspring members were created from each set of parents, and the parameter space consisted of all free parameters.

%

%

\end{document}